\documentclass[aps,twocolumn,showpacs,preprintnumbers,amsmath,amssymb]{revtex4}
\usepackage{colordvi}
\input{epsf}

\def\bc{\begin{center}}
\def\ec{\end{center}}

\def\be{\begin{equation}}
\def\ee{\end{equation}}

 \def\bi{\begin{itemize}}
\def\ei{\end{itemize}}

\def\bea{\begin{eqnarray}}
\def\eea{\end{eqnarray}}

\def\bR{{\sf R}}

 \def\bE{{\bf E}}
\def\bL{{\bf L}}
\def\bC{{\bf C}}
\def\bK{{\sf K}}
\def\bQ{{\sf Q}}
\def\bA{{\sf A}}
\def\bB{{\sf B}}
\def\bW{{\sf W}}
\def\bJ{{\sf J}}
\def\bH{{\sf H}}

 \def\ri{{\rm i}}

  \def\Re {{\it Re}}
  \def\Im{{\it Im}}
  \def\bI{{\bf I}}

\def\wasred{}
\def\wasblue{}

\begin{document}
\title{Mathematical analysis and simulations of the neural circuit for
 locomotion in lamprey}
\author{Li Zhaoping$^1$, Alex Lewis$^1$, and Silvia Scarpetta$^2$}
\affiliation{$^1$University College, London, UK\\
$^2$INFM \& Dept.of Physics, University of Salerno, Italy}
\begin{abstract}
 We analyze  the dynamics of the neural circuit of the
{lamprey} central pattern generator (CPG) 
This analysis provides insights into how neural interactions 
form oscillators and enable spontaneous oscillations in a network 
of damped oscillators, which were not apparent in previous simulations
or abstract phase oscillator models.
We also show how  the different behaviour regimes (characterized by phase and amplitude 
relationships between oscillators) of forward/backward swimming,
and turning, can be controlled using the neural connection
strengths and external inputs. 

\end{abstract}
\maketitle

 \wasred{Locomotion in verbebrates {(walking}, 
 swimming, etc.) is}
generated by  central pattern generators (CPGs) in the spinal cord.
The CPG for  {swimming in lamprey}
is  \wasred{one of the best known
\cite{Grillner2002,Buchanan2001}}, and
has been a model system for 
 \wasred{investigations.}
{It}
produces left-right {anti-phase} oscillatory neural and motor activities
propagating along a body composed of around 100 segments. 
\wasred{A head-to-tail negative or positive oscillation phase
  gradient,
  of about
1\% of an oscillation cycle per segment, gives forward or background 
swimming respectively, and
one wavelength from head to tail.}
 \wasred{External inputs from the brain stem {}{switch} the CPG between 
} forward and backward swimming of
various speeds and turning.
Since {isolated sections} of the spinal cord,
\wasred{ down to 2-3 segments long \cite{Buchanan2001}},
can produce  swimming-like {activity}, 
the
oscillations are thought to be generated by the neurons
within the  \wasred{CPG}.
The neural circuit responsible is shown topologically
in Fig. \ref {fig:circuit}. It has
 \wasred{ipsilaterally projecting 
excitatory ($\bE$) neurons and 
inhibitory ($\bL$) neurons,
and contralaterally projecting 
 inhibitory $(\bC$) neurons, and
provides output to motor neurons via the $\bE$ neurons.}
{All {neurons} project  both intra- and inter-segmentally.
 \wasred{The projection distances are {}{mainly} within a few
segments, especially from $\bE$ and $\bC$ neurons, and  
are longer, and possibly stronger, in the head-to-tail 
or descending direction} \cite{Grillner2002}-\cite{McClellan1998}. 
\begin{figure}[t]
\setlength{\unitlength}{0.6pt}
\begin{center}
\mbox{\epsfxsize=250pt \epsfbox{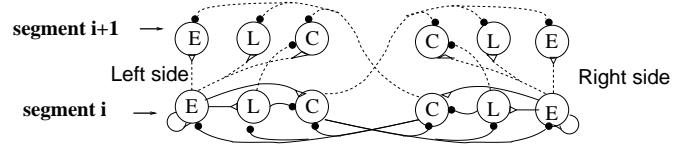}}
\end{center}
\caption{\label {fig:circuit} The lamprey CPG circuit. The solid and dashed lines
denote intra- and inter-segment connections respectively. }
\end{figure}
 Previous analytical work \cite{CohenEtal82, E&K}
{mainly} treated the CPG as a chain of coupled phase oscillators 
 \wasred{in a general form $\dot\theta_i = \omega_i +
 \sum_jf_{ij}(\theta_i,\theta_j)$.
 Here $\theta_i$ is oscillation phase and $\omega_i$
 {is}  intrinsic frequency,
modelling the  
behaviour of  {one} segment, and
$f_{ij}(\theta_i,\theta_j)$
{models} inter-segmental coupling.}
This approach  \wasred{provided important insights into
the conditions {for}
  phase-locked solutions
applicable} to  \wasred{various} systems of coupled oscillators. 
However, 
 \wasred{its generality} obscures the
 {roles of specific} neural types and their 
\wasred{connections}
in generating and controlling behaviour.
\wasblue{More recently, %
{bifurcation} analysis of the dynamics
  of a single segment was carried out,} for a
  phase oscillator model derived from a kinetic (Hodgkin Huxley)
  equation for neurons \cite{Taylor1998}, and   
  for a neural circuit model similar to the one used in this paper
  \cite{Jung}.}
 {Extensive simulations}, including all neural types
   {}{and detailed neural}  properties,
have reproduced many features of experimental data \cite{Grillner2002},
though the {model's} complexity limits further
understanding.

In all previous approaches, it is assumed that a single segment in the
{}{CPG can oscillate} spontaneously, contrary to experimental evidence
that at least 2-3 segments are needed for
oscillations \cite{Buchanan2001}.
We present  \wasred{an} analytical study, confirmed by simulations, of 
a model of the CPG neural circuit in which {}{an isolated} single
segment
has a stable
fixed point, with spontaneous oscillations occurring only  in 
chains of coupled segments.
The phase oscillator approach is not
applicable here since it assumes spontaneously {{oscillating 
individual segments perturbed by inter-segment coupling}.
Including specific cell types and their {connections} enables us to
analyse the role of each of them in generating and controlling
swimming.
We 
show how external inputs select forward
and backward swimming, by {}{controlling
the relative strengths of connections 
between various neurons}, and produces 
turning,  by additional input to one
side of the {CPG} only.
We also {analyse}  behaviour near the {body ends}.

We model the CPG neural circuit 
 \wasred{with} $N = 100$ segments denoted by {$i=1,
.,N$.} 
The vector states 
 \wasred{($\bE_l,\bL_l,\bC_l$) and
 \wasred{$(\bE_r,\bL_r,\bC_r)$}, modelling the membrane potentials of the local populations of neurons  \wasred{at the left and
right side of the body respectively},
with $\bE \wasred{_l} = (E^1 \wasred{_l},E^2 \wasred{_l},\cdots,E^N \wasred{_l})$ etc.,
are modelled as leaky integrators of their inputs:}
 \bea
\dot \bE_{ \wasred{l}} &=& - \bE_{ \wasred{l}} - \bK^0 g_C(\bC_{ \wasred{r}}) + \bJ^0
g_E(\bE_{ \wasred{l}}) +\bI_{E, \wasred{l}}
\nonumber \\
\dot \bL_{ \wasred{l}} &=& - \bL_{ \wasred{l}} - \bA^0 g_C(\bC_{ \wasred{r}}) + \bW^0
g_E(\bE_{ \wasred{l}}) +\bI_{L, \wasred{l}}
{\label {uno}}\\
 \dot \bC_{ \wasred{l}} &=& - \bC_{ \wasred{l}} - \bB^0 g_C(\bC_{ \wasred{r}}) + \bQ^0 g_E(\bE_{ \wasred{l}}) -
\bH^0 g_L(\bL_{ \wasred{l}}) +\bI_{C, \wasred{l}} \nonumber 
 \eea
 with the same equation for 
 swapped subscripts 
 {$(l\leftrightarrow
  r)$.}
{$g_E(\bE_l) = (g_E(E^1_l),\dots,g_E(E^N_l))$}
are the neural activities or firing rates, 
 \wasred{as non-negative (sigmoid-like) activation functions of $\bE_l$,}
and likewise for  \wasred{$g_C(\bC_l)$ and $g_L(\bL_l)$}. $\bK^0$,
$\bJ^0$, 
$\bA^0$, $\bW^0$, $\bB^0$, $\bQ^0$, and $\bH^0$ 
are 
$N\times N$
matrices  \wasred{of} non-negative elements
 \wasred{modeling} the synaptic strengths between neurons.
$\bI_{E, \wasred{l}}$, $\bI_{L, \wasred{l}}$, and $\bI_{C, \wasred{l}}$ are
external inputs, \wasred{including those from the  brain stem},
assumed to be  {static}. 
 \wasred{A left-right symmetric
  fixed point $(\bar\bE,\bar\bL,\bar\bC)$ where
$(\dot\bE,\dot\bL,\dot\bC)=0$ exists by setting
external inputs to $\bI_{E,l} = \bar \bE_l + \bK^0g_C(\bar \bC_{r}) 
-  \bJ^0 g_E(\bar \bE_{l})$ (and analogously for other $\bI$'s).
Dynamics for small deviations from $(\bar\bE,\bar\bL,\bar\bC)$
can be approximated linearly,
{and,} 
with a coordinate rotation
{$(\bE_\pm,\bL_\pm,\bC_\pm) {\equiv }
[(\bE_l,\bL_l,\bC_l) -(\bar\bE,\bar\bL,\bar\bC)]
\pm
[(\bE_r,\bL_r,\bC_r)-(\bar\bE,\bar\bL,\bar\bC)]$,}
transformed into two decoupled modes --
the left-right synchronous mode
$(\bE_+, \bL_+, \bC_+)$ 
and the antiphase mode $(\bE_-, \bL_-, \bC_-)$}.
Swimming requires oscillations, with wavelength of one body length, in
the anti-phase mode while the synchronous mode is damped. The
linearised equations are
\bea
\dot \bE_\pm &=& - \bE_\pm \mp \bK \bC_\pm + \bJ \bE_\pm \nonumber \\
\dot \bL_\pm &=& -\bL_\pm \mp \bA \bC_\pm + \bW \bE_\pm  \nonumber \\
\dot \bC_\pm &=& - \bC_\pm \mp \bB \bC_\pm + \bQ\bE_\pm -\bH
\bL_\pm \label{plus_minus} \eea 
where $\bK \equiv \bK^0 g'_C(\bar \bC )$, 
$\bA \equiv \bA^0 g'_C(\bar \bC )$, $\bB \equiv \bB^0 g'_C(\bar \bC )$, 
$\bJ \equiv \bJ^0 g'_E(\bar \bE )$, 
$\bW \equiv \bW^0 g'_E(\bar \bE )$, $\bQ \equiv \bQ^0 g'_E(\bar \bE )$, and
$\bH \equiv \bH^0 g'_L(\bar \bL)$  
 are effective connection matrices,
\wasred{and
  the $g'(.)$'s denote derivatives}. 
The $\bC$ neurons thus become effectively excitatory in the 
 \wasred{anti-phase mode.}
Noting that the lengths of the neural connections are much
shorter {than} 
the body, 
 \wasred{and that 
isolated sections of spinal cord from any part 
of the body generate oscillations with similar amplitude 
and phase relationships \cite{Grillner2002,Buchanan2001}},
we make the approximation of translation invariance,
so that matrix elements such as $\bJ_{ij}$
depend only on ($i-j$),
and impose the periodic boundary condition, 
{$\bJ_{ij}= \bJ(x)$,} where $x= (i-j) \bmod N$. This is adequate 
\wasred{when
behaviour near body ends is not considered.}
Then all connection matrices {}{commute with each other},
with common eigenvectors (expressed as functions of segment number $x$)
 \wasred{$(\bE(x), \bL(x), \bC(x)) \propto e^{i (2\pi m/N)x}$ for
{integer} eigenmode $-N/2< m \le N/2$.
The system solutions are thus combinations of modes
 $(\bE_\pm (x,t), \bL_\pm (x,t), \bC_\pm (x,t)) \propto
e^{\lambda^\pm _m t + i (2\pi m/N)x}$ where $\lambda^\pm _m$ is
eigenvalue of eq. (\ref{plus_minus})
for mode $m$. Forward swimming results if 
the real part $\Re(\lambda^\pm _m ) < 0$ for all modes except
the antiphase mode with $m=1$, ie $Re(\lambda_1^-)>0$. Then this
mode dominates the 
solution (whose growing amplitude
will be constrained by nonlinearity)
$(\bE (x,t), \bL (x,t), \bC (x,t))
\propto e^{ \Re (\lambda^-_1 ) t -i (\omega t - kx)}$,
with oscillation frequency $\omega \equiv \left|\Im (\lambda^-_1)\right|$ and
wave number $k = 2\pi /N$.
Using the convention $e^{-i\omega t}$ for oscillations, we omitted
the solution $\propto e^{\Re(\lambda^-_1 ) t +i \omega t}$ in 
the conjugate pair of eigenvalues.}
To simplify our system, {we} note from experimental data that in forward swimming,
{$\bE$} and {$\bL$} oscillate roughly in phase within {a}
segment, while  {$\bC$} 
 leads them
 \cite{Buchanan2001}.
We scale our  \wasred{variable definitions}  {so that $\bE_-=\bL_-$
in forward swimming. Then } 
 {eq. (\ref
{plus_minus})} implies 
that 
$(\bK -\bA ) \bC_-  = -(\bJ - \bW ) \bE_-$ in forward swimming.
 \wasred{Since $\bE$ and $\bC$ have much shorter connections than
  the  wavelength of oscillations during swimming, the connection
  matrices have zero elements far from the diagonal, making
  $(\bK -\bA ) \bC_-$ and $(\bJ - \bW ) \bE_-$  roughly either in
  phase or in
  anti-phase 
with $\bC_-$ and $\bE_-$ respectively.
As $\bC_-$ phase leads $\bE_-$, $(\bK -\bA ) \bC_-  = -(\bJ - \bW ) \bE_-$
is impossible unless $(\bJ -\bW)\bE_-=(\bK - \bA)\bC_-=0$.}
For simplicity we henceforth assume $\bJ =\bW$ and $\bK =
\bA$, since non-swimming modes do not concern us. Consequently
$\bE_\pm=\bL_\pm$ and 
\be
\begin{pmatrix}
\dot\bE_\pm \\
\dot\bC_\pm \\
\end{pmatrix}
=
\begin{pmatrix}
\bJ-1 &~~& \mp\bK \\
-(\bH-\bQ) &~~& \mp\bB-1
\end{pmatrix}
\begin{pmatrix}
\bE_\pm \\
\bC_\pm\\
\end{pmatrix}
\label {eqv} \ee 
\wasred{where $\bL$ and $\bE$  {are} treated as a single population 
inhibiting or exciting $\bC$ via connections $\bH-\bQ$.}
{The eigenvalues for mode $m$ are}
 {
\begin{align}
\lambda^+_m &= \left[-2 + J_m - B_m  \pm \sqrt{R_m + 2(B_m^2+J_m^2)}\right]/2 
\nonumber \\
\lambda^-_m &= 
\left[-2 + J_m + B_m  - i \sqrt{R_m }\right]/2.
\label {lams}
\end{align}}
 {$J_m \equiv \sum_x \bJ (x) e^{ -\ri (2\pi m/N) x} $} is the 
eigenvalue of $\bJ$ (and analogously for other matrices),
and {}{$R_m$} is the eigenvalue  {of} 
$\bR\equiv 4\bK  (\bH-\bQ) - (\bB-\bJ)^2$. 

To elucidate {}{the}
conditions {}{needed} for {}{the}  {antiphase} 
{mode with}
$m=\pm 1$ 
for forward or backward swimming to dominate, we {}{analyse the
  bifurcations which occur
  as $\lambda^\pm_m$ for each mode ($m$, $\pm$)
changes as the effective neural connections are varied, either directly or via
the external inputs.}
First, we focus on the left-right mode space 
 {(as in \cite{Taylor1998,Jung} 
for a single segment)}
of $+$ and $-$, i.e., the synchronous 
and antiphase modes,  by simply taking $m =0$.
Then, $J_0$, $B_0$, 
$H_0$, $K_0$, and $Q_0$, 
 are all real and
non-negative, each {}{being} the total connection
strength on a postsynaptic cell from all cells of a particular
type. 
Oscillation in the antiphase mode requires $R_0>0$, necessitating 
$H_0>Q_0$, or that in the AC component of interactions above the background
DC level, $\bC$ neurons receive stronger inhibition from $\bL$ neurons 
than excitation from  $\bE$ neurons. 
Consequently, $\lambda^+_0$ is real and
the synchronous mode is non-oscillatory.
As neural connections increase from zero,
the  {antiphase} 
mode undergoes a Hopf bifurcation 
when  {$\Re (\lambda^-_0)=0$}, 
at {}{$J_0+B_0=2$},
{}{and} the  {synchronous} 
mode undergoes a 
pitchfork bifurcation when {}{$\lambda^+_0=0$, which occurs when
$(B_0+1)(1-J_0) = K_0(H_0 - Q_0)$}.
 {Oscillations result} if
  the Hopf bifurcation has occurred but the
  pitchfork bifurcation has not,
  i.e., $Re(\lambda^-_0)>0> \lambda^+_0$. The condition
  $\lambda^+_0<0$ implies
  $(B_0+1)(1-J_0) > K_0(H_0
  - Q_0)$, necessitating $J_0<1$.   Meanwhile,
  $Re(\lambda^-_0)>\lambda^+_0$
  leads to
  $B_0 > \sqrt {J_0^2 +R_0/2}>J_0$, meaning that there must be
  sufficient  {inhibitory}
connections between left and right $\bC$ cells.
The  $\bJ$, $\bW$, and
$\bQ$  {connections from} $\bE$ cells have to be relatively
  weak,
  consistent with the findings  {of \cite{Jung}}. 
(If $R_0 < 0$, the antiphase mode will undergo {}{a pitchfork}
bifurcation, {}{and the synchronous  {mode} 
  either a pitchfork or Hopf bifurcation}.
 These {}{regimes}  are
 less relevant to modeling the  {lamprey}.)

Assuming the synchronous mode {}{is  damped},
we focus now on  {the} antiphase mode in the $m$ mode space.
{}{Hopf bifurcations occur} sequentially in various  modes $m$ 
in the order of descending $\Re (\lambda^-_m)$. 
Taylor expanding 
$J_m$ (and similarly  {$B_m$, $R_m$}) for small 
wave number $k = 2\pi m/N$ {}{as is relevant
for  {swimming}, 
$J_m = j_0 -i k j_1 - k^2 j_2 + {\mathcal{O} }
(k^3)$
with $ j_n = \sum_x \bJ(x)\frac{x^n}{n!}$}, we have
  \bea
\!2\!\Re {(\!\lambda^-(k))\!} &=& \! - \!2 \! +\! j_0 \! +\!\! b_0\!\!-\!\! k
r_1/(2\sqrt{r_0})\! -\!\! k^2\!
(j_2\!\!+\!b_2)\!\! + \!{\mathcal{O}}\!(k^3\!) \nonumber \, \\
\!2\!\Im {(\!\lambda^-(k))}\! &=& \sqrt{r_0} + {\mathcal{O} } (k) 
\eea
making the mode with
{}{$k \approx -r_1/[4\sqrt{r_0} (j_2+b_2)] $}, {}{which has the largest
 $Re(\lambda^-(k))$,} dominant.
From the definition, {}{$(r_0,j_2, b_2) \ge 0$}, {}{while}
stronger and/or longer connections  in the descending
direction {}{imply $(j_1,b_1)>0$}.
Simply, $\bJ$ (and similarly for other matrices) 
is said to be descending (or ascending), if {}{$j_1>0$ (or  $j_1<0$).}
Hence, {}{if} $\bR$ {}{is ascending}, i.e., {}{$r_1<0$},
$\Re ({\lambda^- }(k))$ {}{increases with} increasing $k$, and the
dominant wave number can be set to $k=2\pi/{N}$ for mode $m=1$
by {tuning} the values of $\bR$, $\bJ$, and $\bB$.
If the connection strengths are such that
only the $m=1$ mode undergoes the Hopf bifurcation, 
forward swimming emerges spontaneously. 
Switching $\bR$ to descending  {leads to} backward swimming.
Note that $\bJ$, $\bB$, $\bH$, $\bK$, and $\bQ$ are all descending, 
  {multiplications} and summations of {}{descending}  connections 
are still {}{descending},  {and} negating a descending
connection makes 
it ascending. 
Since $\bB$ and $\bH$ have to dominate $\bJ$ and $\bQ$ respectively,
$\bR$ is composed of an ascending term $-(\bB-\bJ)^2$ 
and a descending term $4\bK(\bH-\bQ)$. 
{}{Depending on} the relative strengths of {}{these two terms}, 
$\bR $ can be {}{made} ascending or descending to achieve forward
or backward swimming.
This could be achieved by changing the
 {static inputs}
to shift the fixed point $(\bar{\bE},\bar{\bL},\bar{\bC})$ of the 
system to a different gain regime 
$g'_E(\bar \bE), g'_L(\bar \bL), g'_C(\bar \bC)$, and  
thus different effective connection strengths $\bH =
\bH^0g'_L(\bar \bL)$, etc. without
changing the underlying connection structure $\bH^0$.
Alternatively, the  \wasred{external} inputs {might} recruit 
 {extra functional cells} to alter the  {effective} 
connection strengths 
\cite{Kozlov2002}.

When connections are such that additional modes 
satisfy $\Re (\lambda^-_m)>0$, the {}{resulting behavior 
  depends on the nonlinear coupling between modes}.
For illustration, consider nonlinearity
only in $g_C(\bC)$.
 \bea
\dot \bE_{\pm} &=& - \bE_{\pm} \mp  \bK^0 g_\pm(\bC) + \bJ \bE_{\pm} \nonumber \\
\dot \bL_{\pm} &=& -\bL_{\pm} \mp  \bA^0 g_\pm(\bC) + \bW \bE_{\pm}  \nonumber \\
\dot \bC_{\pm} &=& - \bC_{\pm} \mp  \bB^0 g_\pm(\bC) +
\bK'\bE_{\pm} -\bH \bL_{\pm} 
 \eea 
where $g_\pm(\bC) = g_C(\bC_l) \pm
g_C(\bC_r)$. If the nonlinearity is of the form $g_C(x) = x + ax^2 - b
x^3 + \mathcal{O}(x^4)$, {we have} 
  {\begin{align}
g_-(C) &\approx  C_- + a C_+C_- -
bC_-^3/4 -
3bC_-C_+^2/4  \\
g_+(C) &\approx  C_+ + a C_+^2/2 +
aC_-^2/2 -  bC_+^3/4 - 3bC_+C_-^2/4 
\nonumber \end{align}}
Hence, when $C_- = 0$,
$C_+$ cannot excite it since $g_-(\bC) = 0$. 
{However}, if $a\ne 0$, the  \wasred{synchronous mode}, 
will be excited  {passively} 
by the  \wasred{antiphase} mode through the quadratic coupling 
term $aC_-^2/2$, responding with double frequency, as could be easily tested.

To analyse coupling between {}{the}
antiphase modes, we {assume for simplicity that $g_c(\bC)$ is odd, so
  $\bC_+=0$ since the synchronous mode is damped,
  and  $g_-(C_-)=2g_C(C_-/2)$.}
Consider a small perturbation, in the $m'$ mode direction,
to the $m=1$ cycle (the final orbit resulting from a
small deviation from the fixed point in the $m=1$ mode, with
a fundamental harmonic in the $m=1$ mode)
 {such} that
$\bC_-(x) \approx C_1 \cos (2\pi x/N) + C_{m'} \cos(2\pi m' x/N)$ 
with $C_{m'} \ll C_1$.  Expressing 
$g_-(\bC)$ as $g_-(\bC) = \sum_n g_n e^{\ri 2\pi n x/N}$, it can be
shown that {for
  large $N$},
$g_{m'} \approx C_{m'} \bar g'_C$ where $\bar g'_C$ is the derivative
of $g_C$ averaged over the unperturbed cycle.
(More
detailed analysis will be given in a future paper.)
 {Because of the sigmoid form of $g_C(C)$, $\bar g'_C<g'_C(\bar \bC)$.}
Then $C_{m'} \propto e^{\rm\lambda^-_{m'}t}$ 
with $\lambda^-_{m'}$ as in equation (\ref {lams}) except that 
$(B_{m'}, K_{m'})$, values derived from connections from $\bC$ 
cells, are rescaled by a factor  {$\bar g'_C /g'_C(\bar \bC)<1$.}
{}{Thus the swimming cycle at large amplitude always
  remains stable against
  perturbation in other modes, even when the fixed point is unstable
  against these perturbations}.

{}{
  If $Re(\lambda_1^-) \gg Re(\lambda_{m'}^-) \gtrsim 0$,
  the $m'$ cycle  will have a small amplitude
and hence $\bar g'(C) \sim g'(\bar{C})$ and it will be unstable
against perturbation in the $m=1$ mode. For larger
$Re(\lambda_{m'}^-)$
the amplitude of the cycle is larger and either cycle will be stable.
} Suppose  the neural connections are such that {}{ the $m=\pm 1$
  cycles, giving forward or backward swimming,  are
  both stable}. {}{The system would then display hysteresis, with
  the final behaviour depending on the initial conditions.} 
{}{Forward or backward swimming could then be selected by}  
transient  inputs from the brain stem,
rather than by 
 {setting constant inputs}
  as described above.
This {}{seems less likely to be} the actual selection mechanism since
experiments on fictive swimming  
(presumably with random initial conditions) 
seldom observe  {spontaneous} backward swimming. {}{However, the
  forward swimming could simply have
  a larger basin of attraction than backward
  swimming.} 

When the lamprey turns,  neural activities  {on left and right sides are
unequal.} This is realizable by adding an additional
constant  {input}  to one side in the animal and
in models \cite{McClellanHagevik1997, Kozlov2002}, leading to {}{unequal
mean}  {activities} 
without
disrupting the oscillations, 
provided that the gains $g'(.)$ are
roughly constant  \wasred{near} the fixed points.
 {}{Simulations (Fig.2) confirm the
analysis above.}

\begin{figure}[!!!hhhhhhhh!]
\setlength{\unitlength}{1.0pt}
\begin{center}
\begin{picture}(350, 170)
\put(0,80){{\epsfxsize=240pt \epsfbox{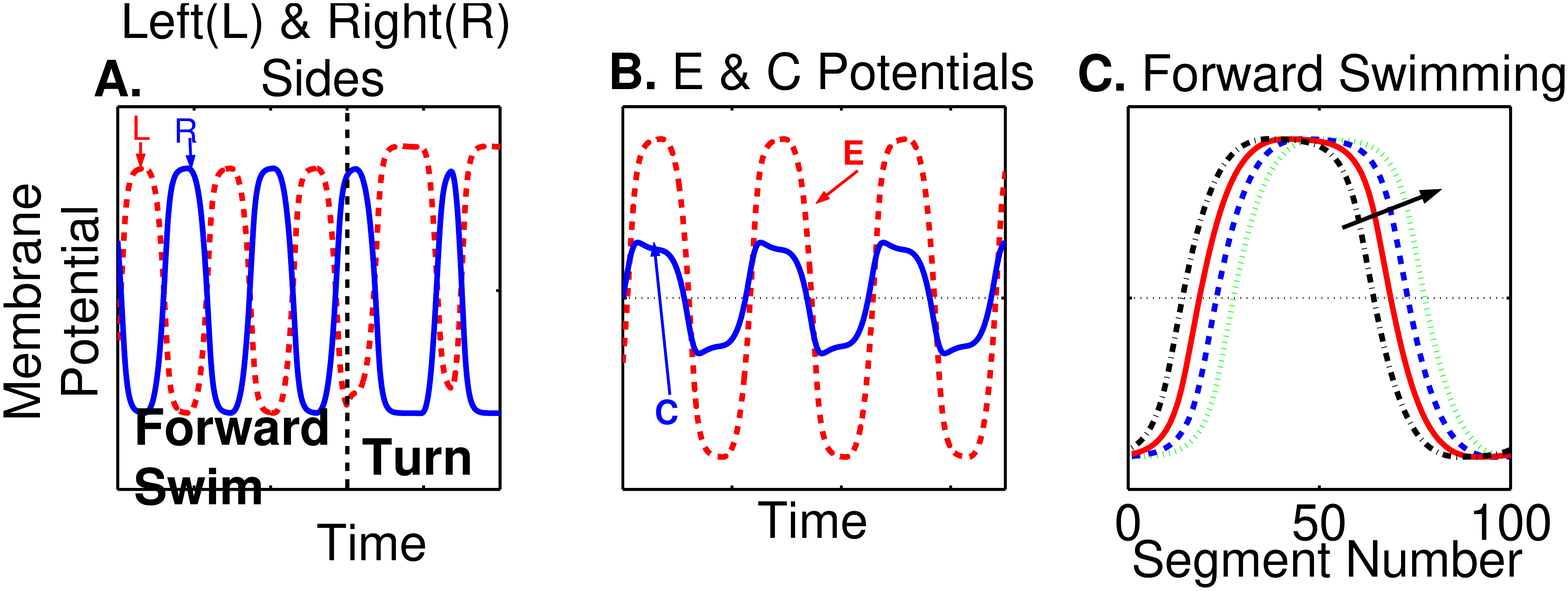}}}
\put(0,-15){{\epsfxsize=240pt \epsfbox{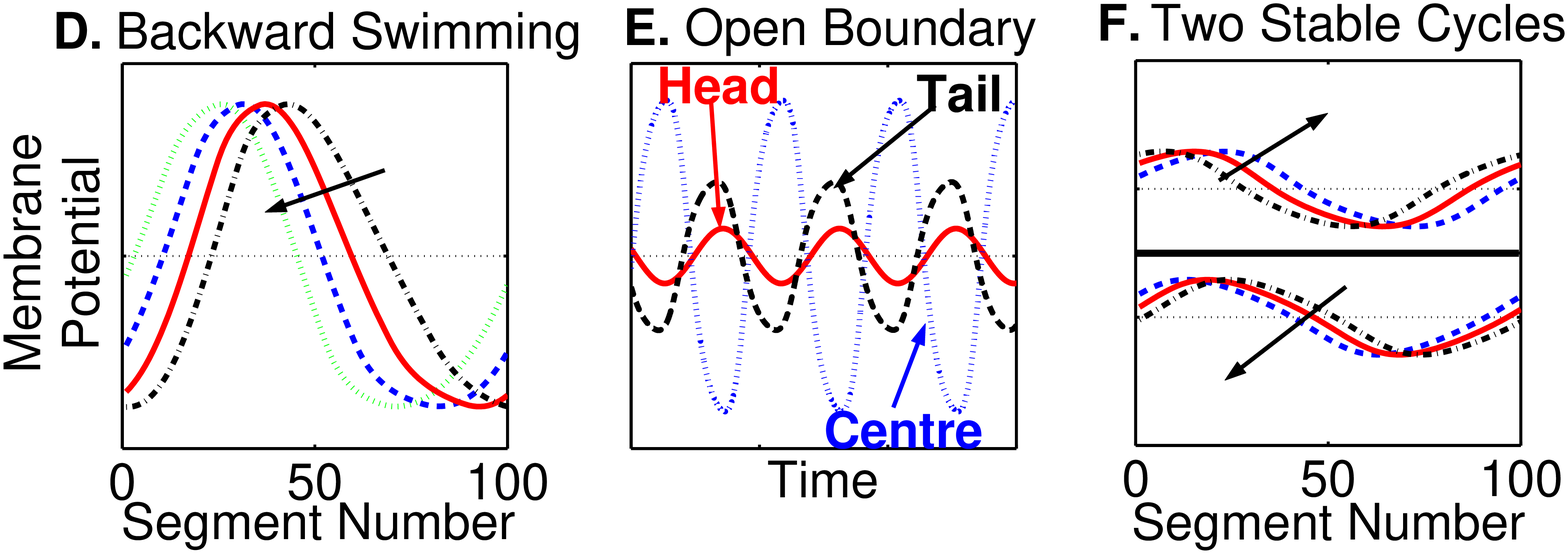}}}
\end{picture}
\end{center}
\caption{Simulations. A: Membrane potentials of $\bE$
population on  {either} side of one segment during forward
swimming and turning. The oscillations are in anti-phase between
the two sides. Turning is induced by an additional constant input
to  one side only,  {starting} at the time indicated by the dashed line. B: 
$\bC$ slightly phase leads $\bE$ during forward swimming. 
C \& D:  {Waveform} of $\bE$ along the body 
in forward and backward swimming, at consecutive times increasing 
in the direction indicated
by the arrows, in the translational invariant model.
The switch from forward to backward swimming is achieved by increasing 
the strength of $\bH$ and $\bQ$.
E: Oscillation waveforms (note different amplitudes) in body
segments at head, tail and centre of the body,  {without} 
translational invariance. F: Forward and backward swimming from
  different initial conditions, with the same connection strengths and inputs.
}
\end{figure}

To study behaviour  {at the body ends}
or in {}{short sections of spinal cord \cite{Grillner2002}}, or
equivalently to see the effects of longer connections,
we abandon 
{}{translation invariance}.
  {}{Eliminating 
$\bC$ in eq. (\ref {eqv})}, 
the minus mode has:
 \be
 \ddot \bE + (2-\bJ -\bB) \dot \bE + [1-\bJ - \bB +\bB\bJ + \bK(
\bH-\bQ)]\bE = 0, \nonumber
 \ee
or, oscillator $i$ is driven by  \wasred{force $F_i$} from {other} oscillators
\be
\begin{split}
\ddot \bE_i &+ (2-\bJ_{ii}-\bB_{ii}) \dot\bE_{i} + [1-
\tilde\bR_{ii}] \bE_{i} 
\label {osci_couple} \\
        &= F_i \equiv \textstyle{\sum_{j \ne i} F_{ij} \equiv \sum_{j \ne i}
(\bJ_{ij} +
\bB_{ij}) \dot \bE_j +\sum_{j\ne i} \tilde\bR_{ij} \bE_j}
 \end{split}\nonumber\ee 
where $\tilde\bR = \bB+\bJ -\bB\bJ
-\bK(\bH-\bQ)$. 
 {}{The intrinsic oscillation},
$\bE_i \sim e^{\lambda t}$, 
{}{is damped, $\Re(\lambda ) = -1 +(\bJ+\bB)_{ii}/2
 <0$},
as 
indicated by  {experiments} \cite{Grillner2002,Buchanan2001}.
We estimate $F_i$ 
 {using} the approximation that oscillators $j\ne i$ still
{behave as}  $\bE_j \propto e^{-{\rm i }(\omega t -kj)}$. 
{}{We then have} 
$F_i = \alpha_i  \dot \bE_i + \beta _i\bE_i$, where
\begin{align}
\alpha_i &\equiv 
 \textstyle{\sum_{j\ne i} \left[(\bJ + \bB)_{ij}  \cos (k(j-i))
  -\tilde\bR_{ij} \sin (k(j-i)) /\omega \right]}
 \nonumber \\
\beta _i  &\equiv 
 \textstyle{\sum_{j\ne i} \left[(\bJ + \bB)_{ij} \omega \sin (k(j-i))
      + \tilde\bR_{ij} \cos (k(j-i))\right]}
\nonumber \end{align}
The term $\alpha _i\dot \bE_i$ when $\alpha_i>0$ feeds oscillation
energy into the {$i^{th}$ (receiving)} oscillator,
\wasred{causing {}{emergent} oscillations
in {}{ coupled damped} oscillators.}
We divide {}{$\alpha_i =\alpha_{i,\text{desc}}+\alpha_{i,\text{asc}}$}
  into the {}{descending and ascending}
  parts, with summations over  {}{$\sum_{j<i}$
and $\sum_{j>i}$} respectively.
Hence, for $i=1$, $\alpha_1 = \alpha_{i,\text{asc}}$; for  $i=N$,
$\alpha_N = \alpha_{i,\text{desc}}$, {and for $1\ll i \ll N$},
$\alpha_{i} = \alpha_{1} + \alpha_{N}$. Since the first
and last {}{segments}
oscillate due to the driving force from other
oscillators, $\alpha_{1}>0$ and $\alpha_{N}>0$. Consequently,
$\alpha_{1}< \alpha_{N/2}$ and $\alpha_{N}< \alpha_{N/2}$.
{}{Further}, since {}{descending connections
  are stronger},  \wasred{it is most
likely that, for $1\ll i \ll N$},  $\alpha_{i,\text{desc}}>
\alpha_{i,\text{asc}}$. Consequently, $\alpha _{1} < \alpha _{N}$.
{Hence, the rostral oscillator has a smaller amplitude than the caudal one,
  which in turn has a smaller amplitude than the central one
   {(Fig 2(E))}. 
Firing rate saturation and variations of the fixed point along the
body may obscure {}{this  pattern} in experimental data,
although body movements are indeed smallest near the head \cite{Paggett1998}.
\wasred{Similarly, oscillation amplitudes {}{will}
  be reduced in sections of 
spinal cords shorter than the typical lengths of inter-segment
connections, and
will eventually be zero in ever shorter sections, 
as observed in experiments \cite{Buchanan2001}.}

In summary,  {analysis of} a model  of the
CPG neural circuit in lampreys  {has given}
{}{new insights into} the neural connection structures {needed}
to
generate and control the swimming behaviour.
In particular, we predict that the contra-lateral connections between $\bC$
must be stronger than the self-excitatory connection strength of
the $\bE$ neurons; that the
$\bC$ neurons are more inhibited
(in their AC components) by $\bL$ neurons than excited by
the $\bE $ neurons;
and {have shown} how different swimming regimes can be selected by
scaling the strengths of the various neural connections without changing 
 \wasred{the} connection patterns.
Our framework should help to provide further
insights into CPGs of animal locomotion. 

{\bf Acknowledgements}: 
We thank Peter Dayan for  \wasred{comments,  Carl van Vreeswijk for
  {}{help with literature}, and
the Gatsby Charitable Foundation for support.}

\end{document}